\documentclass[final,5p,authoryear,twocolumn]{elsarticle}

\usepackage{graphicx}
\usepackage{epsfig}
\usepackage{amssymb}

\journal{Planetary and Space Science}

\begin{document}

\begin{frontmatter}


 \title{Dispersive MHD turbulence in one dimension} 

\author{D. Laveder}
\ead{laveder@oca.eu}
\author{L. Marradi}
\ead{marradi@oca.eu}
\author{T. Passot}
\ead{passot@oca.eu}
\author{P.L. Sulem}
\ead{sulem@oca.eu}
\address {Universit\'e de Nice Sophia Antipolis, CNRS,  Observatoire de la C\^ote d'Azur,  \\
BP 4229, 06304 Nice Cedex 4, France}  



\begin{abstract}
Numerical simulations of dispersive turbulence in magnetized plasmas based on the Hall-MHD description are presented, 
assuming spatial variations along a unique direction making a prescribed angle with the ambient magnetic field.
Main observations concern the energy transfers among the different scales
and the various types of MHD waves, together with the conditions for the establishment of pressure-balanced structures.
For parallel propagation, Alfv\'en-wave transfer to small scales
is strongly inhibited and rather feeds magnetosonic modes, unless
the effect of dispersion is strong enough at the energy injection scale.
In oblique directions, the dominantly compressible character of the turbulence is pointed out
with, for quasi-transverse propagation, the presence of conspicuous kinetic Alfv\'en waves.
Preliminary simulations of a Landau fluid model incorporating relevant linear kinetic effects  
reveal the development of a significant plasma temperature anisotropy leading to recurrent instabilities.
\end{abstract}

\end{frontmatter}

\section{Introduction}

Turbulence in magnetized plasmas remains a main issue in the understanding of
the dynamics of media such as the solar corona, the interstellar
medium, the solar wind or the planet magnetosheaths. In the solar wind for example
the turbulent cascade extends much beyond the ion Larmor radius. One of the questions concerns
the spectrum of the magnetic fluctuations that  displays  a power-law behavior on a broad
range of wavenumbers,  with a conspicuous change of slope  near  the inverse ion
gyroradius \citep{LSNMW98,GR99,AMM06,SGRK09}. This effect is often  associated with the
influence of wave dispersion, induced by the Hall current 
\citep{GSRG96,Gal06,ACVS06,GB07,SCPVS07,ShSh09},
but could also result from a superposition of cascades of kinetic Alfv\'en waves and
ion entropy fluctuations,  as suggested by studies based on the gyrokinetic formalism
\citep{HDC08,HCD08,SCD09}.

At scales large compared with the
ion inertial length or the ion Larmor radius, the usual MHD description
provides a satisfactory description of regimes where, due to the
presence of a strong ambient field, a dominant effect is
the anisotropic energy transfer to Fourier modes with large transverse wavenumbers
(see e.g. \citet{GG97,OM05} and references therein). This suggests that
the  dynamics of transverse small scales may be amenable to
a reduced MHD description (\citep{ZM92} and references therein),
possibly including Hall current \citep{GMD08} or,
when retaining scales significantly smaller than the ion Larmor radius,
to a  gyrokinetic approach \citep{HCD06,SCD09}.
The latter that appears to be very efficient in describing
strongly-magnetized near-equilibrium fusion plasmas is still under discussion concerning its
applicability to space and astrophysical plasmas \citep{MSD08}. In the solar wind 
for example  magnetic fluctuations may be comparable to the ambient field.
Furthermore, longitudinal transfer could a priori be non  negligible 
in a compressible regime,  at scales
where  Hall current  and  kinetic effects play a significant role.
A weak turbulence theory performed on the Vlasov-Maxwell system
was recently developed \citep{YF08}, showing the existence of a
parallel cascade of low-frequency Alfv\'en waves through a
three-wave decay process mediated by ion-sound turbulence, in a regime where wave-particle
interactions are neglected.
Addressing this issue by direct numerical simulations of the Vlasov-Maxwell
equations being still difficult on the present-day computers,
the question arises whether a similar cascade
can be observed within a fluid model that retains important ingredients of the above
theory, such as compressibility and dispersion.
As a first step, we address  the  problem
within the simplest description provided by Hall-MHD (HMHD)
with Ohmic and viscous dissipations, together with a  large-scale external
driving acting on the transverse components of the velocity or magnetic field.
We specifically concentrate on a  one-dimensional setting  where the variations of the
fields are restricted to a direction making a prescribed angle with the ambient magnetic field,
a framework that already reveals a manifold of complex dynamical processes that deserve detailed
investigations before including additional physical and multidimensional effects.
In the case of quasi-transverse propagation,
we also present simulations of a model that extends the HMHD by retaining pressure anisotropy,
Landau damping and finite Larmor
radius effects up to transverse scales significantly smaller than the ion Larmor radius. This
approach developed in \citet{PS07} extends the so-called Landau fluid model
initiated in  \citet{SHD97} for the MHD scales where Landau damping is the only relevant
kinetic effect.

The paper is organized as follows. Section 2 briefly reviews the Hall-MHD description
and its one-dimensional reduction.  Section 3  concentrates on the case where the dynamics
takes place in the direction of the ambient field. The case of oblique propagation is
addressed in Section 4. Landau fluid simulations retaining small-scale kinetic
effects are reported in Section 5. Our conclusions are summarized in Section 6.

\section{The Hall-MHD description}

HMHD can be viewed as a bi-fluid description of a plasma, 
where electron inertia is neglected. The presence of the Hall term in the 
generalized Ohm's law allows a decoupling 
of the ion fluid from the electron one in which the magnetic field lines are frozen.
The  validity conditions of HMHD are discussed in 
\citet{H09} where comparisons with kinetic theory are presented.
Choosing  as units  the Alfv\'en speed, the 
amplitude of the ambient magnetic field, the equilibrium density and 
the ion inertial length $l_i$ (defined as the ratio of the 
Alfv\'en speed to the ion gyrofrequency), the HMHD equations (for the ion fluid) read
\begin{eqnarray}
&&\partial _{t}\rho + \mbox{\boldmath $\nabla$} .( \rho{\bf v} )  =
0\label{eq:mhd1}\\ 
&&\rho ( \partial _{t}{\bf v} + {\bf v}. \mbox{\boldmath $\nabla$} {\bf v} )  =    
- \frac{\beta}{\gamma} \mbox{\boldmath $\nabla$}\rho ^{\gamma} + ( \mbox{\boldmath $\nabla$}
 \times {\bf b} ) \times {\bf b}  \label{eq:mhd2} \\
&&\partial_{t}{\bf b} -  \mbox{\boldmath $\nabla$}\times ( {\bf v}\times {\bf b})  =  
 - \mbox{\boldmath $\nabla$}\times \Big ( \frac{1}{\rho} (\mbox{\boldmath $\nabla$}
\times {\bf b} ) \times {\bf b} \Big ) \label{eq:mhd3} \\
&& \mbox{\boldmath $\nabla$}.{\bf b}  =  0, \label{eq:mhd4} 
\end{eqnarray}
where the total $\beta$ parameter is the square ratio of the sound to Alfv\'en velocities, and 
a  polytropic equation of state $p \propto  \rho^\gamma$ is  assumed for both ions and electrons.

When the spatial variation is restricted  to a dependency on the $x$ coordinate
along a direction making an angle $\theta$ with the ambient magnetic field ${\bf B}_0=(\cos \theta, 
\sin \theta,0)$, one gets
\begin{eqnarray}
&&\partial_{t}\rho + \partial_{x}(\rho v_x) = 0\label{eq:mhd1d_1}\\ 
&&\partial_{t} v_x + v_x \partial_{x} v_x = 
- \frac{1}{\rho} \partial_{x} \Big( \frac{\beta}{\gamma} \rho^{\gamma}
+\frac{1}{2} (b_{y}^{2} + b_{z}^{2}) \Big) \nonumber \\
&& \quad + \frac{\mu_x}{\rho}\partial_{xx} v_x 
\label{eq:mhd1d_2}\\
&&\partial_{t} v_{[y,z]} + v_x \partial_{x} v_{[y,z]}
= \frac{\cos \theta}{\rho } \partial_{x} b_{[y,z]} \nonumber \\
&& \quad  + \frac{\mu_{[y,z]}}{\rho}\partial_{xx} v_{[y,z]} + f_{[y,z]}^v\label{eq:mhd1d_3}\\
&&\partial_{t} b_{[y,z]} - b_x\partial_{x} v_{[y,z]} + \partial_{x}(v_x b_{[y,z]})
=  \nonumber \\
&&\quad \pm \cos \theta \partial_{x} \Big( \frac{1}{\rho} \partial_{x} b_{[z,y]} \Big) 
+ \kappa_{[y,z]} \partial_{xx} b_{[y,z]} + f^b_{[y,z]}, \label{eq:mhd1d_5}
\end{eqnarray}
where driving  and dissipation have been supplemented in both the velocity and magnetic
field equations. Here, 
the subscript $[y,z]$ refers to the vector component along the $y$ or the $z$ direction,
or to the value of the viscosity acting on the corresponding velocity component. 
The $\pm$ sign in front of the Hall term depends on the considered component of the magnetic field.
No (artificial) hyperviscosity and magnetic diffusivity nor spectral filtering are 
used in the simulations. Instead, anisotropic dissipations are assumed. 
In the case of parallel propagation, different viscosities and diffusivities
are taken  in the directions parallel
and transverse to the ambient field, by prescribing  $\kappa_y =\kappa_z = \mu_y=\mu_z \ll \mu_x$.
For  oblique propagation,  we assume  smaller coefficients in the direction 
perpendicular to the plane defined by the magnetic field and the direction of propagation, in the form
$\kappa_z = \mu_z \ll \mu_x=\mu_y= \kappa_y$.

The driving is assumed to act either on the velocity (kinetic driving)
or the magnetic field (magnetic driving) components. 
For parallel propagation, we
prescribe  $f^v_y=f^v_z$ or $f^b_y=f^b_z$  while, for oblique propagation, the driving reduces to 
$f^v_z$ or to $f^b_z$.  Such a driving is  supposed to minimize 
the  sonic components, as it is acting 
on field components perpendicular to the ambient field in parallel propagation 
and to the plane defined by the ambient field and the propagation direction when the latter is oblique.
The values of the diffusivity and viscosities in the various 
directions are chosen as the minimal values 
(depending on the spatial resolution and of the physical parameters of the runs) needed to  
accurately resolve all the retained scales.

In all the simulations, we take $\gamma = 5/3$ and  $\beta=2$.
Each component of the kinetic or magnetic driving (generically denoted $f$) is a white noise in time
defined by its Fourier transform $\widehat{f}_k = C \xi \sqrt{F_k /\Delta t}$
where $\xi$ is a Gaussian 
random variable with zero mean and unit variance, chosen independently at each time step.
This ensures a constant mean flux of energy injection that can be 
chosen at will,  as in the usual phenomenology of the turbulent cascades.
Furthermore, such a driving process avoids an artificial enhancement of a specific type of waves and 
enables the emergence of the dominant modes as the result of the nonlinear dynamics.
The spectral distribution $F_k = k^4 \exp( -(2k^2/k_f^2))$ is peaked about a wavenumber $k_f$.

The HMHD system is integrated in a periodic domain using a 
Fourier pseudo-spectral method where most of the aliasing is removed  
by spectral truncation of the computed nonlinear terms at $2/3$ of the maximal wavenumber.
The spatial resolutions given in the following sections are the effective ones, 
after aliasing has been suppressed.
In all the simulations, the temporal scheme is a third-order low-storage Runge-Kutta \citep{Wil80}.
Resolving all the temporal scales present in the system, this scheme accurately 
preserves the dispersion relation of all the linear modes retained in the simulation, in contrast
with implicit or semi-implicit schemes \citep{LBPS09}.

For convenience, we collected in Table 1, the main parameters characterizing the simulations 
discussed in the forthcoming sections.

\begin{table*}
\begin{center}
\begin{tabular}{|c|c|c|c|}
  \hline
  run & domain size  & propagation angle & driving \\
  \hline
  A & $L=16 \pi$  & $\theta= 0^\circ$ & {\rm kinetic}, $C=0.1$, $k_f l_i=1/2$\\
  B & $L=16 \pi$  & $\theta= 0^\circ$ & {\rm magnetic}, $C=0.1$, $k_f l_i=1/2$\\
  C & $L=4 \pi $  & $\theta= 0^\circ$ & {\rm kinetic}, $C=6.25 \times 10^{-3}$, $k_f l_i=2$\\
  D & $L=4 \pi $  & $\theta= 0^\circ$ & {\rm magnetic}, $C=6.25 \times 10^{-3}$, $k_f l_i=2$\\
  E & $L=16 \pi$  & $\theta= 45^\circ$ & {\rm kinetic}, $C=0.1$, $k_f l_i=1/2$\\
  F & $L=4 \pi $  & $\theta= 45^\circ$ & {\rm kinetic}, $C=6.25 \times 10^{-3}$, $k_f l_i=2$\\
  G & $L=4 \pi $  & $\theta= 45^\circ$ & {\rm magnetic}, $C=6.25 \times 10^{-3}$, $k_f l_i=2$\\
  H & $L=16 \pi$  & $\theta= 80^\circ$ & {\rm kinetic}, $C=0.1$, $k_f l_i=1/2$\\
\hline
\end{tabular}
\caption{Simulation parameters of HMHD simulations}
\end{center}
\end{table*}

As seen in the following, in spite of the turbulent regime achieved in the
HMHD simulations discussed in this paper, signatures of the  linear waves are often 
present. It is thus useful to briefly review the linear theory 
of eigenmodes for the HMHD equations  in the absence of dissipation and driving. 
 By linearizing 
eqs. (\ref{eq:mhd1d_1})-(\ref{eq:mhd1d_5}) about the equilibrium state associated to
$\rho=1$, $b_x=\cos \theta$, $b_y=\sin \theta$, one derives that the 
(real) eigenfrequencies $\omega_i$ with 
$i=1,2,3$ obey the dispersion relation
\begin{eqnarray}
&&\omega^6 -k^2 (1 + \beta + \cos^2 \theta + k^2  \cos^2 \theta) \omega^4 \nonumber \\
&& + k^4  \cos^2  \theta (2 \beta + 1 + \beta k^2) \omega^2 - 
\beta k^6  \cos^4  \theta = 0, \label{disprela}
\end{eqnarray} 
where it is sufficient to concentrate on the positive solutions.

\begin{figure}[b]
\centerline{
\includegraphics[height=12cm,width=8cm]{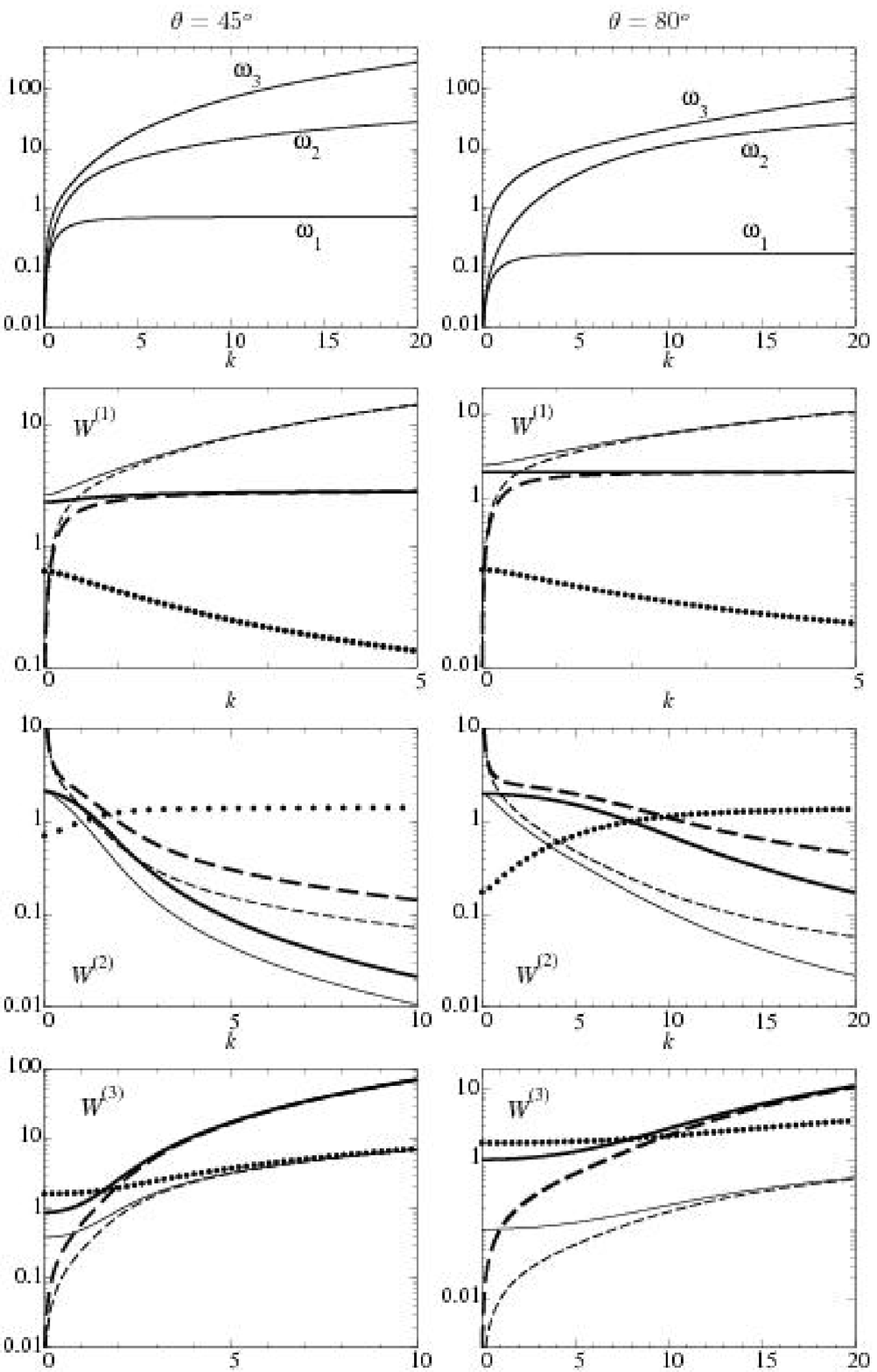}
}
\caption{Wavenumber dependency of the linear HMHD eigenfrequencies (top) and eigenmode components
${\bf W}^{(1)}$, ${\bf W}^{(2)}$, ${\bf W}^{(3)}$ (from middle to bottom), 
when prescribing $\rho=1$,
for propagation angles $\theta = 45^0$ (left) and $\theta = 80^0$ (right): 
magnetic field $b_y$ (thick solid line) and $b_z$
(thick dashed line); velocity components $v_y$ (thin solid line) and $v_z$ (thin dashed line);
velocity component $v_x$ (dotted line).}
\label{figure_linearmodes}
\end{figure}

Assuming an oblique propagation, the corresponding eigenmodes are given by 
\begin{eqnarray}
&&{\bf W}^{(i)} = (\rho, v_x = \alpha_{v_x}^{(i)}\rho, v_y = \alpha_{v_y}^{(i)} \rho, 
 v_z = \alpha_{v_z}^{(i)} \rho, \nonumber \\
&& b_y = \alpha_{b_y}^{(i)} \rho, 
b_z = \alpha_{b_z}^{(i)} \rho)
\end{eqnarray}
with 
\begin{eqnarray}
&&\alpha_{v_x}^{(i)}=\frac{\omega_i}{k}  \\
&&\alpha_{b_y}^{(i)}=\frac{(k^2 \cos^2 \theta - \omega_i^2) \sin \theta }
{ k^4 \cos^2 \theta - \Big (\frac{k^2}{\omega_i}\cos^2 \theta - \omega_i\Big ) ^2 }\\
&&\alpha_{b_z}^{(i)}=\frac{-i k^2\omega_i \sin \theta \cos \theta }
{ k^4 \cos^2 \theta - \Big (\frac{k^2}{\omega_i}\cos^2 \theta - \omega_i\Big ) ^2 }\\
&&\alpha_{v_y}^{(i)}= -\cos \theta \frac{k}{\omega_i} \alpha_{b_y}^{(i)}\\
&&\alpha_{v_z}^{(i)}=  -\cos \theta \frac{k}{\omega_i} \alpha_{b_z}^{(i)}.
\end{eqnarray}
The variation of these coefficients with the wavenumber $k$ is displayed in fig. 
\ref{figure_linearmodes} for $\theta = 45^o$ and $\theta = 80^o$.

For parallel propagation ($\theta=0$), the sonic wave for which $\omega =\sqrt{\beta} k$
decouples from the eigenmodes corresponding to (circularly-polarized)  
Alfv\'en waves. The latter obey 
$b_y + i b_z  = -(\omega/k)(v_y + i v_z)  \propto \exp [ i \sigma (kx - \omega t) ] $ 
where $\sigma=-1$ or $1$, depending of the left-hand 
(LH) or right-hand (RH) polarization, with the  dispersion relation
$\displaystyle{\omega_{\sigma} = (\sigma/2) k^2 + k (1+ k^2/4)^{1/2}}$.
In the  large wavenumber limit, the frequency of LH polarized waves 
saturates, while it grows like $k^2$ in the case
of RH polarization. At frequencies beyond the HMHD asymptotics,
the RH polarized Alfv\'en waves are continued into whistler waves, 
while the LH polarized waves become ion-cyclotron waves. 
It is noticeable that the dispersive parallel Alfv\'en waves are also exact solutions 
of the nonlinear HMHD problem.

For an oblique direction of propagation and $\beta >1$, 
we still refer to the eigenmode corresponding to the smallest eigenvalue $\omega_1$, 
that is dominated by the field components $v_y$, $v_z$ (fig. \ref{figure_linearmodes}),
as the ion-cyclotron wave.
In the case of quasi-transverse propagation, it is often called the kinetic Alfv\'en wave.  
The intermediate mode corresponding to the eigenvalue $\omega_2$ is the one that 
reduces to a pure sonic mode for $\theta=0$. 
For oblique propagation it is dominated by the $v_x$ component at small scales only, 
while at large scales $b_z$ and $v_z$ prevail.
The mode at highest frequency $\omega_3$, which corresponds 
to the whistler wave for $\theta=0$, has properties opposite to those of the intermediate
one. For oblique propagation, it is usually viewed as a whistler wave at small scales only \citep{KV94}. 
For the sake of simplicity, we will keep this terminology at large scales also, where this mode 
is strongly compressible.

\section{Parallel dynamics}

\subsection{Large-scale driving}

We first consider a domain of extension 
$L= 16\pi$ in a direction parallel to the ambient magnetic field. We 
use a transverse kinetic or a magnetic driving 
characterized by the parameters $C=0.1$ and  $k_f=0.5$
corresponding to a  mode index $n_f=4$. 
Defining the ion inertial wavenumber as $k_i= 2\pi/l_i$,
this corresponds to a ratio $k_i/k_f = 4 \pi$.  
A spatial resolution of $1024$ grid points is used, which prescribes a time step 
as small as $\Delta t=5\cdot 10^{-5}$, due to the dispersion relation of the whistler modes.  
We use a viscosity $\mu_x= 10^{-1}$ in the equation for the parallel velocity,
and equal viscosity and magnetic diffusivity $\mu_y =\mu_z  = \kappa_y=  \kappa_y = 10^{-6}$ 
for the transverse fields. This relatively large viscosity  in the parallel direction is required
because of the development of shock waves.

\begin{figure}
\centerline{
\includegraphics[height=3.7cm,width=8cm]{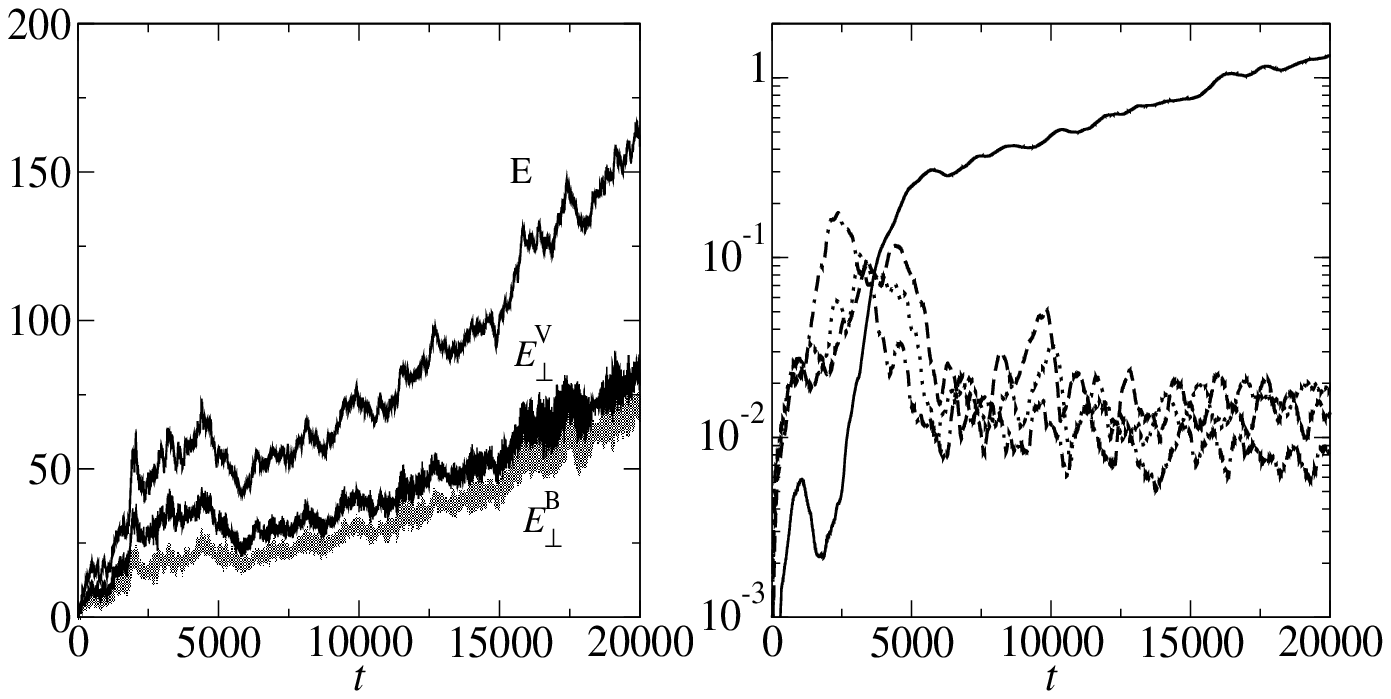}
}
\caption{Left: time evolution of the total energy and of the transverse kinetic
and magnetic contributions for  run A with kinetic forcing discussed in Section 3.1. 
Right: time evolution of the magnetic energy 
$|\widehat{\bf b}_y|^2+|\widehat{\bf b}_z|^2$ for the Fourier modes of index 1(solid line), 2(dashed), 
3(dotted), 4(dash-dotted), for the same simulation.}
\label{energy_par_largebox}
\end{figure}

In the case of kinetic driving (run A), 
fig. \ref{energy_par_largebox} (left) shows the time evolution
of the total energy of the system (from which the initial value has been subtracted) 
$E= \int (\frac{\rho}{2} |{\bf v}|^2 + \frac{1}{2}(|{\bf b}|^2-1) + \frac{\beta}{\gamma(\gamma-1)} 
(\rho^\gamma-1)) dx$. It also displays the contributions of the transverse
kinetic $E_\perp^V=\int \frac{\rho}{2} |{\bf v}_\perp|^2dx$ and magnetic 
$E_\perp^M=\int\frac{1}{2}|{\bf b}_\perp|^2dx$ energies  that turn out to be  comparable
and much larger than the parallel kinetic and the internal energies (not shown).
It is conspicuous that the energy does not saturate, even after an integration  time $t=20000$.
Figure \ref{energy_par_largebox} (right) displays the time evolution of the magnetic energy 
on the  modes of index $1,2,3,4$ (a similar behavior is obtained for the kinetic modes).
We observe that after a while, the mode of index $n=1$ becomes strongly dominant, 
indicating a significant inverse transfer to larger and larger scales
(leading to a non saturation of the energy),  as expected from an equilibrium
thermodynamical argument \citep{SMC08}, up to the moment when it reaches the  
size of the computational domain.

\begin{figure}
\centerline{
\includegraphics[height=3.7cm,width=8cm]{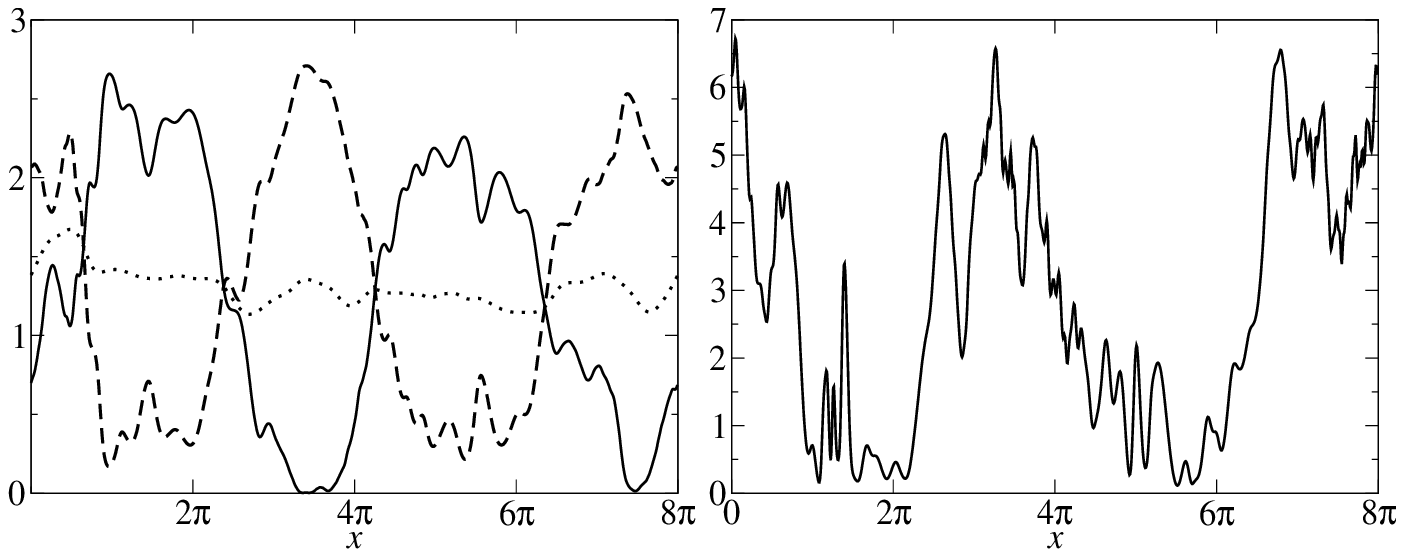}
}
\caption{Left: profile of the  the magnetic pressure $|{\bf b}_\perp|^2/2$ (solid line)
and of the thermal pressure $\beta \rho^{\gamma}/\gamma$ (dashed line), 
together with their half-sum (dotted line) 
showing pressure balance at time $t=20000$ for the run A. 
Right: transverse velocity field intensity $|{\bf v}_\perp|^2$ at the same time.} 
\label{fields_pb_par_largebox}
\end{figure}

\begin{figure}
\centerline{
\includegraphics[height=3.7cm,width=8cm]{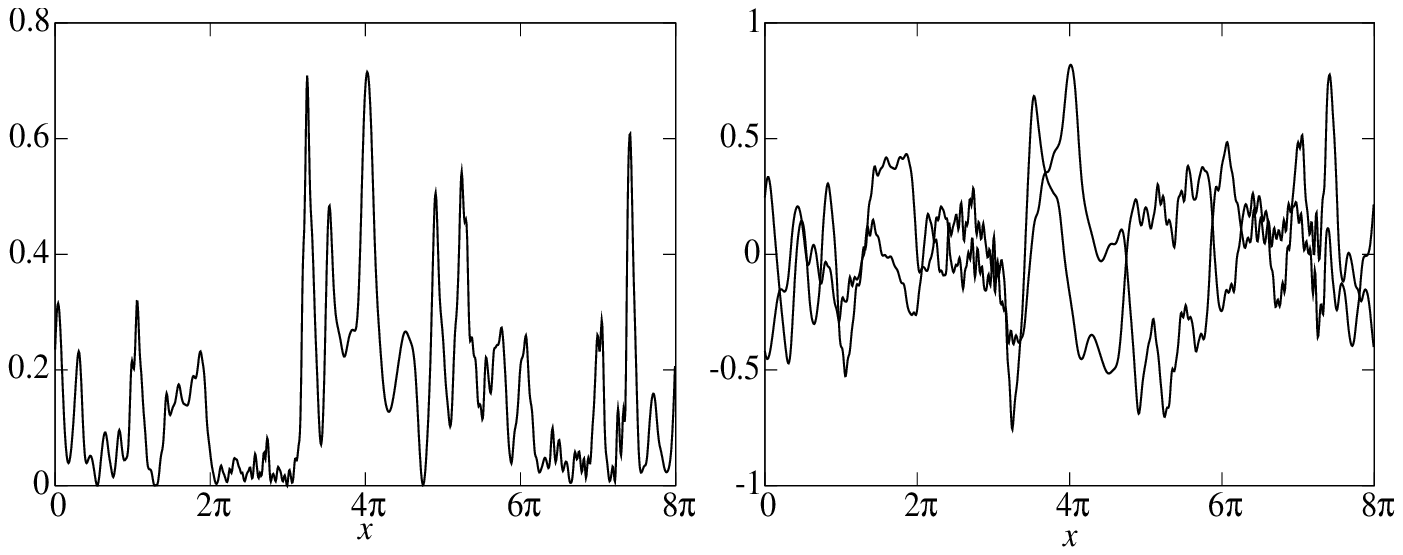}
}
\caption{Transverse velocity field intensity ($|{\bf v}_\perp|^2$) (left) and individual components $v_y$, $v_z$ 
at time $t=18200$ for  run A, 
after filtering the mode of index $n=1$.}
\label{fields_par_largebox}
\end{figure}

A main observation is the establishment of a pressure-balanced state (fig. \ref{fields_pb_par_largebox}, left)
that is not limited to the largest scales, in the sense that pressure balance is still observed when the 
largest Fourier modes are filtered out. Inspection of the square transverse velocity  $|{\bf v}_\perp|^2$
(fig. \ref{fields_pb_par_largebox}, right) shows the presence of small scales superimposed to the $n=1$ mode,
which are more conspicuous than on the magnetic component $|{\bf b}_\perp|^2$ displayed on the left panel.  
When filtering out the modes of index $n=1$, the square transverse velocity $|{\bf v}_\perp|^2$ 
reveals the presence of solitonic waves which in many  instances  survive collisions  
and preserve their coherence on several time units  (fig. \ref{fields_par_largebox}, left).
 It is of interest to notice that dispersive pressure-balanced structures are commonly observed in 
space plasmas (see e.g. \citep{SSG03}).
The right panel that displays the individual transverse velocity components also shows the presence of
rotational discontinuities (near $x=4\pi$), together with the existence of fluctuations at very small scales. 

\begin{figure}
\centerline{
\includegraphics[height=3.7cm,width=8cm]{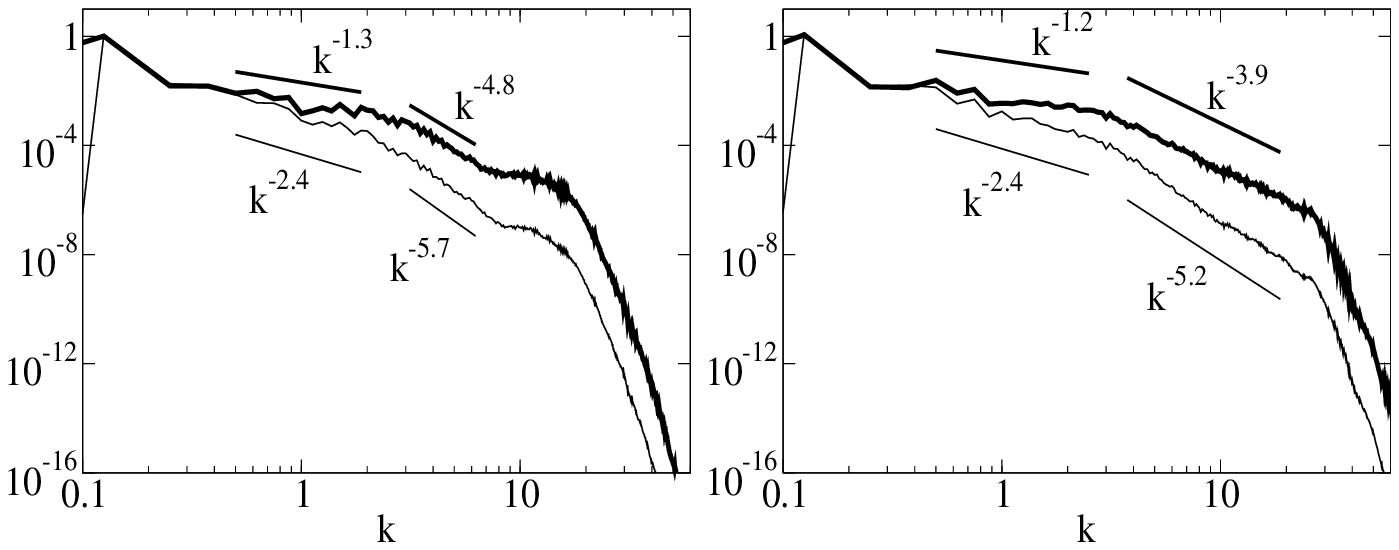}
}
\caption{Kinetic $|\widehat{\bf v}_y|^2+|\widehat{\bf v}_z|^2$ (thick line)
and magnetic $|\widehat{\bf b}_y|^2+|\widehat{\bf b}_z|^2$ (thin line) spectra
averaged over the time interval $t=18200-18250$ (left), and $t=19500-19550$ (right)
for run A. 
}
\label{spectra_par_largebox}
\end{figure}

Figure \ref{spectra_par_largebox} displays the transverse kinetic  
$|\widehat{\bf v}_y|^2+|\widehat{\bf v}_z|^2$ (thick line)
and magnetic $|\widehat{\bf b}_y|^2+|\widehat{\bf b}_z|^2$ (thin line) spectra,
averaged over the time intervals  $t=18200-18250$ (left), and $t=19500-19550$ (right). 
The left panel corresponds to a stage of the evolution  for which fig. \ref{fields_par_largebox} 
provides  instantaneous snapshots. 
We can distinguish various spectral ranges, and in particular 
two different power-law domains at large and intermediate scales whose  exponents fluctuate in time while 
preserving a kinetic spectrum shallower than  the magnetic one. 
Such a dominance of the kinetic on the magnetic contribution suggests an ion-cyclotron turbulence
since for these waves the transverse velocity and magnetic field components are in a ratio $k/\omega > 1$.
The transition between the two spectral ranges occurs near $k \approx k_i=2 \pi$ 
(reminiscent of observations in the solar wind and magnetosheath \citep{LSNMW98,GR99,AMM06,SGRK09}) 
that corresponds to the typical scale of the solitonic structures seen in fig. \ref{fields_par_largebox}. 
The flat region  visible at larger wavenumbers is the spectral signature of the small-scale fluctuations  
visible in physical space. 
At longer times (right panel), the intermediate power-law range has extended as the soliton amplitude increased,
taking over the flat spectral region.

In the case of a magnetic driving (run B), a similar overall dynamics is obtained but on a longer time scales
(not shown). The energy accumulation on the largest scale is slower by a factor four to five,
and the pressure balance is less conspicuous even  at  the end of the simulation 
($t= 10 000$). Furthermore, no clear power law spectra have developed. 

\begin{figure}
\centerline{
\includegraphics[height=3.7cm,width=8cm]{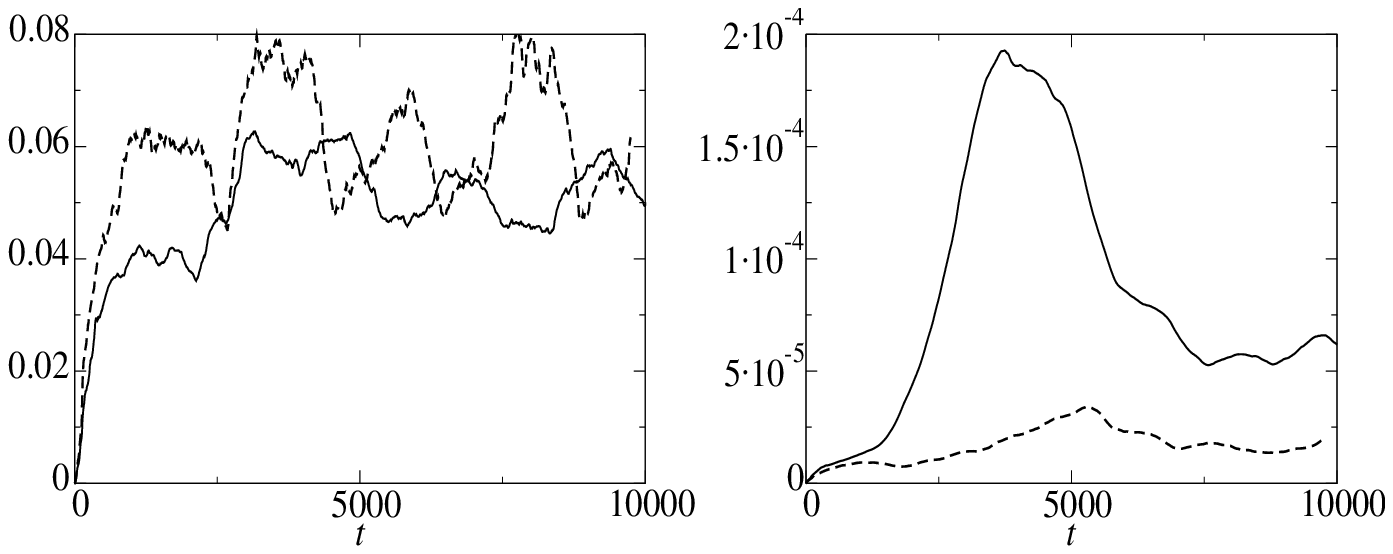}
}
\caption{Left: time evolution of the parallel dissipation $D_\|$
 for the kinetic (solid line) 
and magnetic (dashed line) drivings (runs A and B respectively) described in Section 3.1. 
Right: same for the perpendicular dissipation $D_\perp$.}
\label{diss_par_largebox}
\end{figure}

Further insight on the dynamics is provided by the dissipation of the various 
fields. Figure  \ref{diss_par_largebox} (left) displays 
the time evolution  (averaged over a time interval of 1000 time units) 
of the viscous dissipation $D_\|= \int \mu_x (\partial_x v_x)^2 dx $ 
originating from the parallel velocity
for the two types of driving. The dissipation observed is slightly larger in the case of magnetic driving 
and reflects a stronger  direct energy transfer to  sonic waves. 
This observation is consistent
with the better pressure balance obtained with kinetic than with magnetic forcing. 
On the right panel  is displayed the (similarly averaged) dissipation 
$D_\perp=\int (\mu_y(\partial_x v_y)^2  + \mu_z(\partial_x v_z)^2
+ \kappa_y (\partial_x b_y)^2  +  \kappa_z(\partial_x b_z)^2) dx $ of the transverse  fields, 
that turns out to be  larger in the case of kinetic driving, indicating
a more efficient transfer to small scales for the transverse fields. 

\begin{figure}
\centerline{
\includegraphics[height=3.7cm,width=8cm]{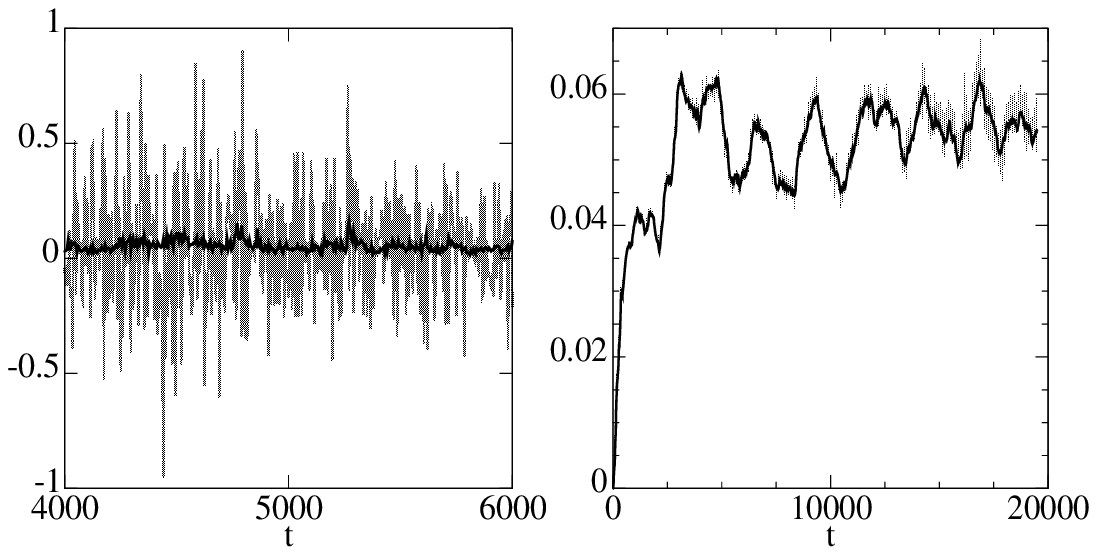}}
\caption{Time evolution of the parallel dissipation (dark line) 
and of the rate of energy transfer $S$ from the Alfv\'en to magnetosonic
waves (light line) for run A.   
Left: instantaneous quantities. Right: time-averaged quantities over $\Delta t_{\rm ave}=1000$.}
\label{enbalance_par_largebox}
\end{figure}

Energy dissipation affects dominantly the parallel velocity component,
the transverse dissipation being  about two orders of magnitude smaller.
This  suggests that the injected energy is mainly transferred to 
sonic waves and dissipated through  a cascade of acoustic waves. For parallel propagation,
writing the equations for the energy density of the Alfv\'en and acoustic waves in the form
($\mu_\perp \equiv \mu_y = \mu_z, \kappa_\perp \equiv \kappa_y = \kappa_z$)
\begin{eqnarray}
&& \partial_t\Big ( \frac{\rho}{2} |{\bf v}_\perp|^2 + \frac{1}{2}|{\bf b}_\perp|^2\Big )\nonumber \\
&& + \partial_x\Big (\frac{1}{2} \rho |{\bf v}_\perp|^2v_x +\frac{1}{2} \rho |{\bf b_\perp|^2}v_x -
{\bf v}_\perp\cdot {\bf b}_\perp + h_x \Big ) = \nonumber\\
&&-\frac{1}{2} |b_\perp|^2\partial_x v_x + \mu_\perp {\bf v}_\perp \cdot \partial_{xx}{\bf v}_\perp + 
\kappa_\perp {\bf b}_\perp \cdot \partial_{xx}{\bf b}_\perp \nonumber \\
&& + {\bf v}_\perp \cdot {\bf f}_\perp^v
+ {\bf b}_\perp \cdot {\bf f}_\perp^b\\
&&\partial_t \Big (\frac{\beta}{\gamma(\gamma-1)} \rho^\gamma + \frac{\rho}{2} v_x^2 \Big)\nonumber \\
&& +\partial_x \Big (\frac{\beta}{\gamma -1} p + \frac{1}{2} \rho v_x^3 
+  \frac{1}{2} v_x |{\bf b}_\perp|^2\Big ) \nonumber\\
&& = \frac{1}{2} |b_\perp|^2 \partial_x v_x +  \mu_x v_x  \partial_{xx}v_x +  v_x f_x^v +  b_x f_x^b 
\end{eqnarray}
where 
\begin{equation}
h_x= \frac{1}{\rho} \epsilon_{j1q} b_jb_x \partial_x b_q
\end{equation}
arises from the Hall term, we get that the energy transfer from the Alfv\'en to sonic waves 
is given by 
\begin{equation}
S= \frac{1}{2}\int |{\bf b_\perp}|^2 \partial_x v_x dx.
\end{equation}
Figure \ref{enbalance_par_largebox}
displays in the case of kinetic driving
the time variation of $S$,  together with that of the parallel (viscous) dissipation.
Left panel, which corresponds to instantaneous quantities, shows that the 
amplitude fluctuations of the transfer are much larger than those of the dissipation.
We nevertheless observe on the right panel which displays the same quantities
averaged on a time interval of 1000 time units, that the parallel dissipation identifies 
in the mean with the energy transfer. A similar behavior is visible with the  magnetic driving.

\begin{figure}
\centerline{
\includegraphics[height=3.5cm,width=8cm]{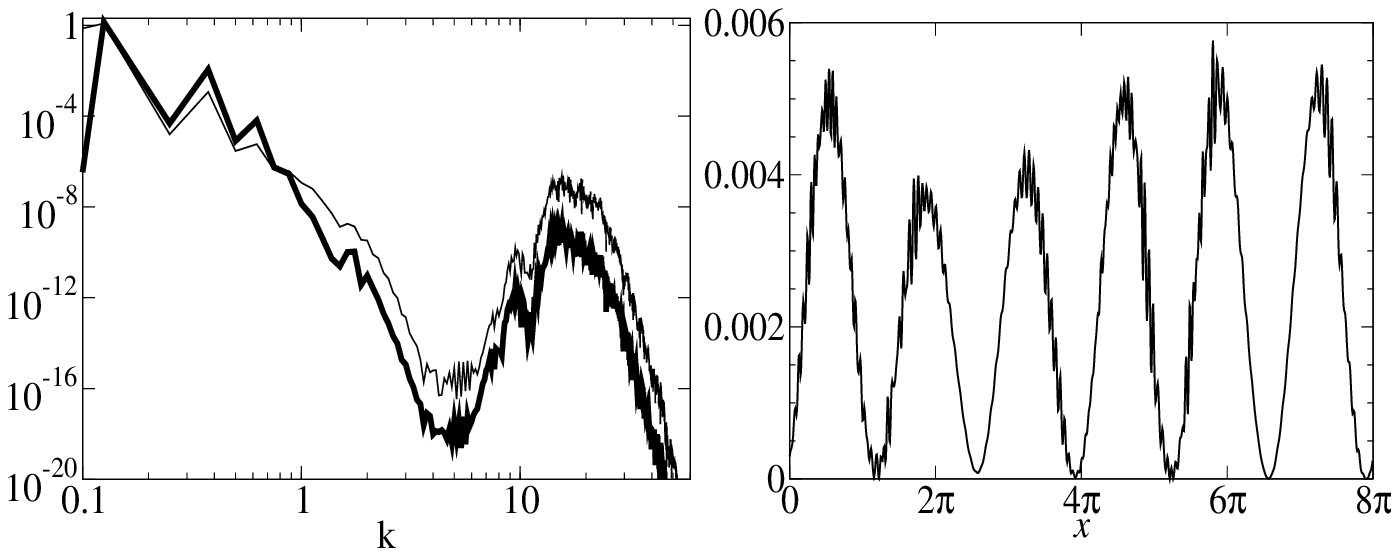}
}
\caption{Left: kinetic  $|\widehat{\bf v}_y|^2+|\widehat{\bf v}_z|^2$ (thick line)
and magnetic $|\widehat{\bf b}_y|^2+|\widehat{\bf b}_z|^2$ (thin line) spectra
averaged over the time interval $t=28000-28050$ for the kinetically driven run of Section 3.1
with monitored forcing. Right: transverse velocity field intensity 
($|{\bf v}_\perp|^2$) at time $t=28000$, in the same conditions.}
\label{spectra_par_largebox_ec}
\end{figure}

The above observations indicate that sonic wave turbulence is the dominant phenomenon, 
although small 
scales also form on the transverse field components  on a longer time scale. 
In order to address more precisely the nature of the transfers, it is of interest to perform a 
simulation where the final time $t=20 000$ of the previous simulation is taken as 
initial condition and the energy is no longer injected at a constant rate, 
but the driving is monitored in order to maintain the total energy almost constant.
This constraint leads to a drastic change on the transverse spectra which,
within a few thousands of time units, loose energy
at the scale of the solitons displayed in fig. \ref{fields_par_largebox}, 
leading to a conspicuous spectral gap (fig. \ref{spectra_par_largebox_ec}, left).
In physical space, solitonic structures have indeed disappeared, 
the profile of $|{\bf v}_{\perp}|^2$ reducing to small-scale oscillations superimposed to 
large-scale quasi-sinusoidal structures, as seen in 
fig. \ref{spectra_par_largebox_ec} (right) where the strongly dominant mode $n=1$ 
has been filtered out. 
At scales larger than the spectral gap, the pressure-balanced state persists, while at smaller scales 
viscous and Ohmic dissipations are negligible and dispersion is significant, thus permitting the small scales to persist. 
This suggests that, in the present simulation, the transfer of Alfvenic energy
from large to intermediate Alfv\'en scales
which in the previous simulation maintained the solitons,
does not take place in the absence of a sufficiently strong driving,
and thus cannot compensate the transfer of energy from the Alfv\'enic to the sonic modes
at the soliton scales.
The dynamics thus turns out to be significantly different from ordinary turbulence
and should rather be viewed as a structure-dominated regime displaying significantly less
universal properties.
 
\begin{figure*}[p]
\centerline{
\includegraphics[height=20cm,width=16cm]{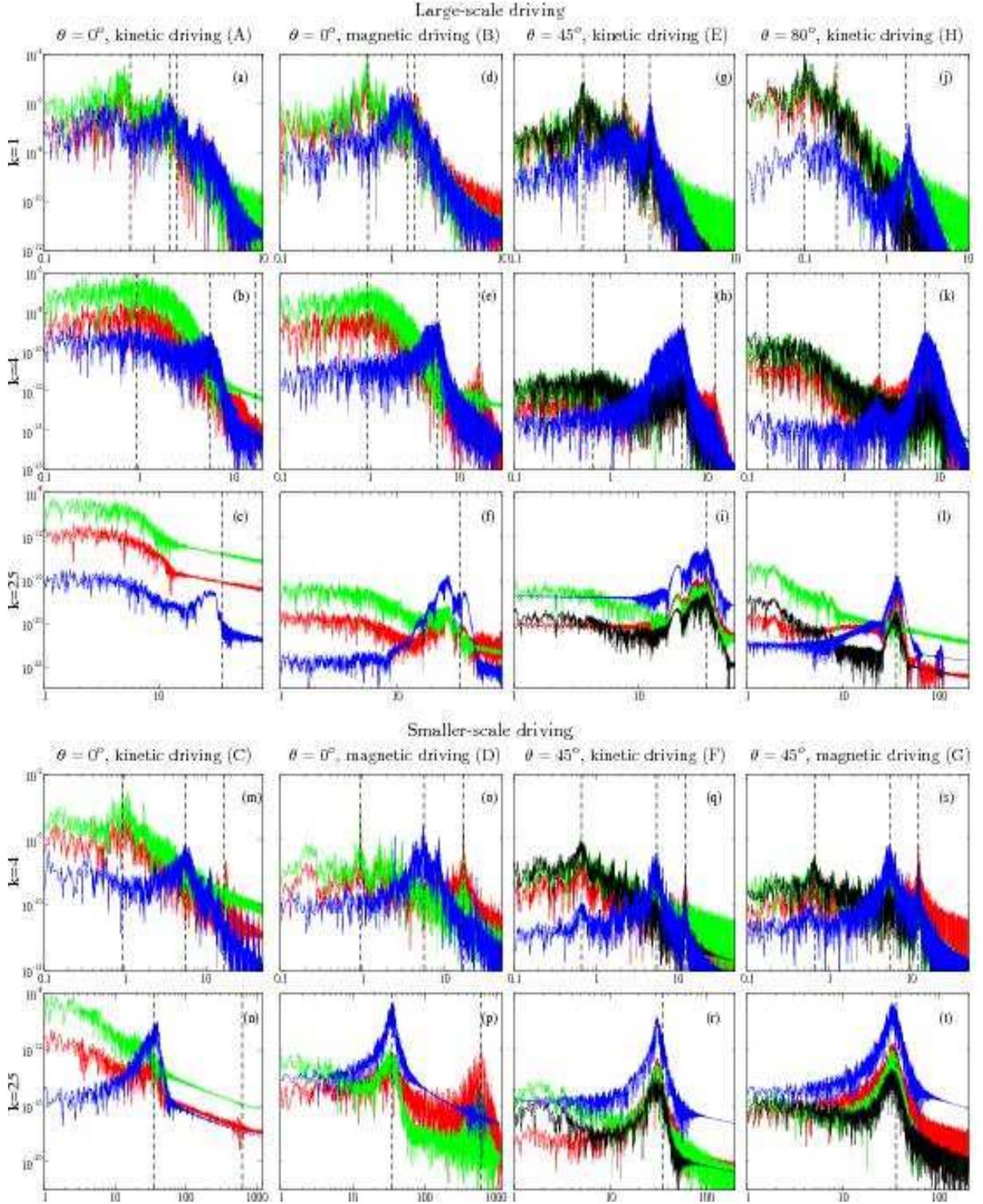}
}
\caption{Frequency spectra $|\widehat{b}_z|^2(k,\omega)$ (red), 
$|\widehat{v}_z|^2(k,\omega)$ (green), $|\widehat{v}_y|^2(k,\omega)$ (black), 
$|\widehat{v}_x|^2(k,\omega)$ (blue) as  functions of $\omega$,
at various wavenumbers $k$ for the runs described in Sections 3 and 4 (see Table 1).
The vertical lines refer to the three linear eigenfrequencies given by the dispersion relation
(\ref{disprela}). When only two lines are visible, they correspond to the two higher frequencies, 
while when only one is present, it is the intermediate one.}
\label{figure_color}
\end{figure*}

A further characterization of the dynamics is provided by
a frequency analysis on the spatial Fourier modes of the fields components. 
In contrast with the incompressible regime where the presence of the sole
Alfv\'en waves makes convenient the analysis of time records at given points in physical 
space \citep{DM09}, 
the possible dominance of different modes at different scales in the present case, 
leads to discriminate between different Fourier modes.
Figure \ref{figure_color}
displays the temporal spectrum of the  spatial Fourier modes $k=1$, $k=4$ and $k=25$
respectively, for  the components $v_x$ (blue), $b_z$ (red) and $v_z$ (green)
for kinetic (run A: panels a,b,c) and magnetic (run B: panels d,e,f) driving.
The analysis has been performed during a time interval close to the end of the simulations.
At large scale ($k=1=2k_f$), two peaks are visible. 
They  correspond, up to a slight infrared shift possibly due to 
the presence of large-scale coherent structures,
to the ion-cyclotron ($\omega \approx  0.62$) and the sonic  ($\omega\approx 1.4$) frequencies.
These peaks are wider with kinetic than with magnetic forcing, 
indicating stronger nonlinear couplings in the former case which  
is also characterized by a larger amount of transverse energy at all scales, 
with a ratio between the kinetic  and magnetic components in qualitative
agreement  with  the linear ion-cyclotron eigenmode.
At smaller scales ($k=25 \approx 4k_i$), the relative importance 
of the sonic peak decreases in the case of kinetic driving, 
but remains significant for magnetic driving, indicating 
a higher level of sonic turbulence in the latter regime. 

\subsection{Driving at smaller scales}

In a second series of simulations, the global effect of the Hall term
has been increased by injecting energy at smaller scales, with the aim
to investigate the influence of dispersion on the energy transfers. 
To address this issue, the extension of the computational domain
was reduced to $L = 4 \pi$ and the system driven at $k_f = 2$, 
leading to a ratio $k_i/k_f=\pi$. The rate of energy injection was lowered
by the same factor as the domain extension by taking $C=6.25\cdot 10^{-3}$, 
in order to preserve the same amount of energy injection per unit length. 
Reducing the resolution to $256$ mesh  points, 
ensures the same maximal wavenumber as in the simulations of Section 3.1.  
To preserve numerical stability, we had to use time steps 
$\Delta t=3.125 \cdot 10^{-6}$ and $\Delta t=2.5 \cdot 10^{-5}$ for the kinetic and magnetic 
drivings respectively. The required viscosities and magnetic diffusivities
are now $\mu_x= 10^{-2}$, $\mu_y= \mu_z = \kappa_y= \kappa_z  = 10^{-5}$ in both cases.

\begin{figure}
\centerline{
\includegraphics[height=3.7cm,width=8cm]{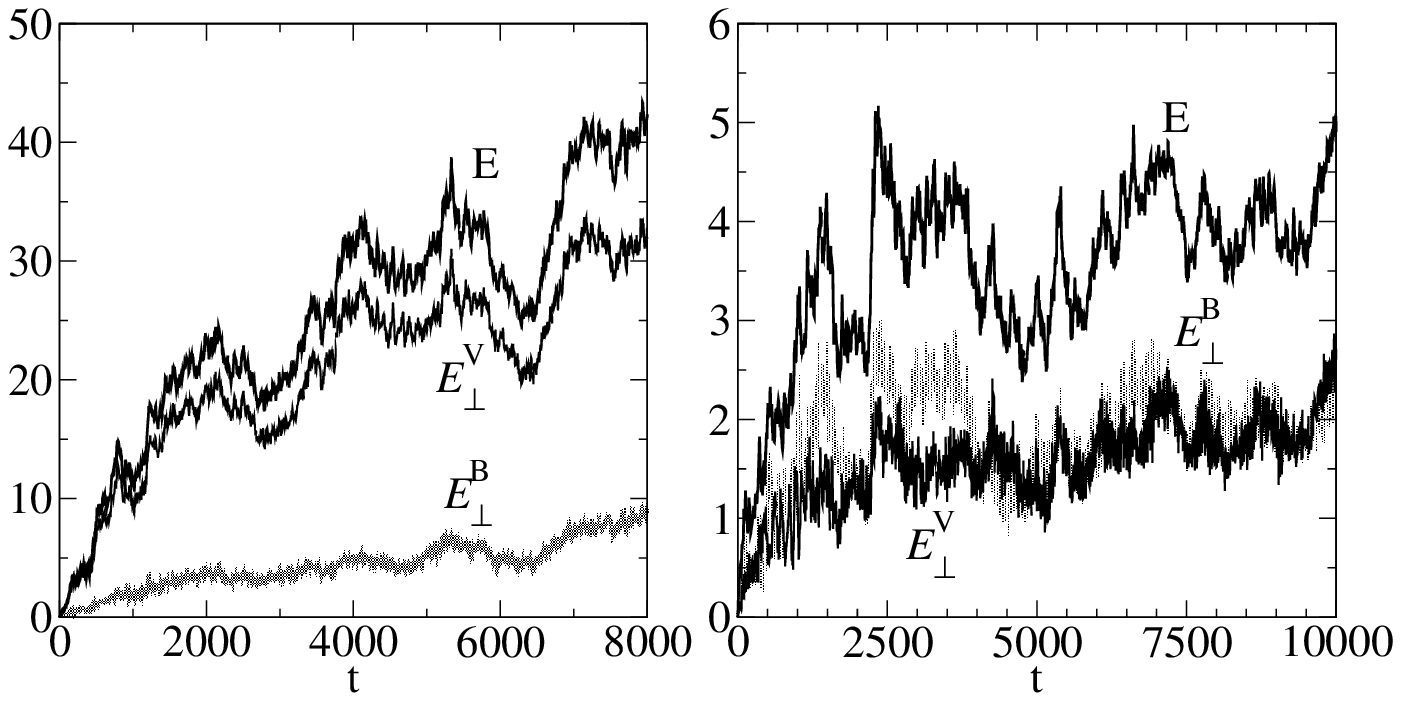}
}
\caption{Time evolution of the total kinetic and magnetic
energies for kinetic (left) or magnetic (right) drivings (runs C and D respectively)
in the conditions described in Section 3.2.}
\label{energy_par_smallbox}
\end{figure}

\begin{figure}
\centerline{
\includegraphics[height=3.7cm,width=8cm]{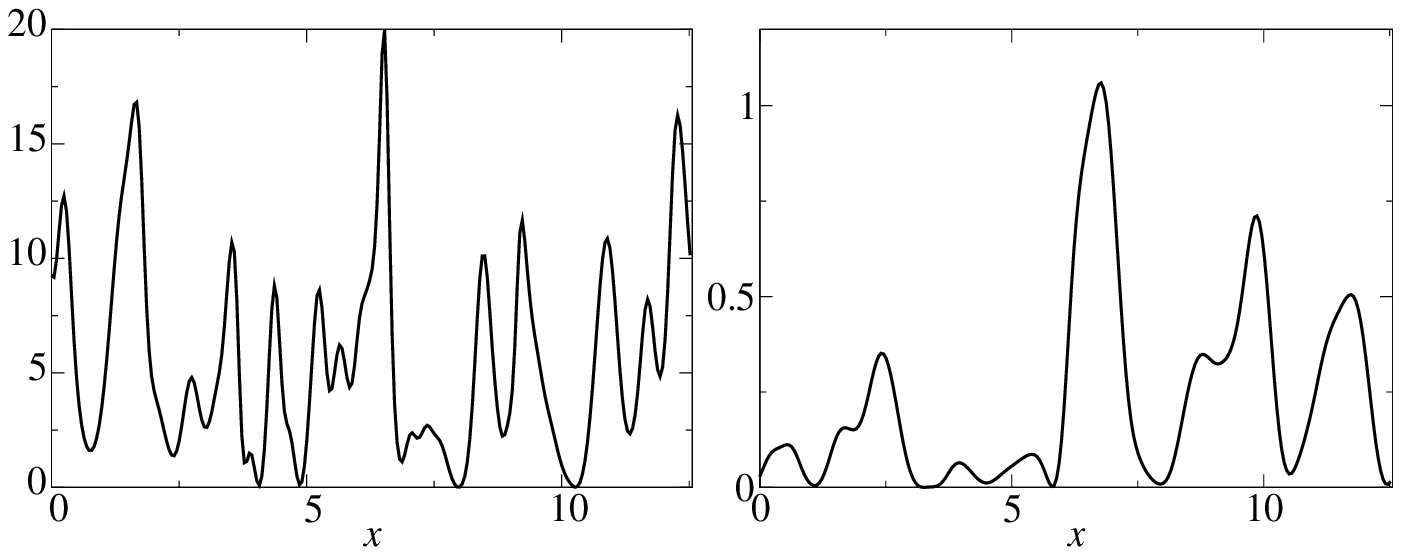}
}
\caption{Profile of $|{\bf v}_{\perp}|^2$ at $t=4000$ for kinetic (run C, left) 
and magnetic (run D, right) driving for the run described in Section 3.2.}
\label{fields_par_smallbox}
\end{figure}

Kinetic (run C) and magnetic (run D)  drivings lead now to significantly different dynamics.
In the former case, the inverse cascade is less efficient than in the 
conditions of Section 3.1. The energy of the modes with $n=1$ still increases 
up to the end of the simulation (not shown), but  after a time of about 7000 times units,
the accuracy of the simulation deteriorates, the temporal resolution becoming insufficient. 
The total energy grows slowly but in contrast with the conditions of Section 3.1,  
the transverse magnetic  energy is now 
significantly lower than the kinetic one (fig. \ref{energy_par_smallbox}, left).
Differently, in the case of a magnetic
driving, kinetic and magnetic energies are  comparable, and the system reaches a stationary
steady state where the  energies saturate  (Fig. \ref{energy_par_smallbox}, right).
In the case of a kinetic forcing, pressure-balanced structures are still present, while
with a magnetic driving, no pressure balance establishes, 

Figure \ref{fields_par_smallbox} displays the profiles of $|{\bf v}_\perp|^2$ for 
kinetic (left panel) and magnetic (right panel) drivings, showing that more small 
scales develop in the former case. The structures  
are traveling much more slowly than with the large-scale forcing. 
Furthermore, when comparing transverse velocity and magnetic fields, we note that the 
latter is significantly smoother for both types of drivings.

\begin{figure}
\centerline{
\includegraphics[height=3.7cm,width=8cm]{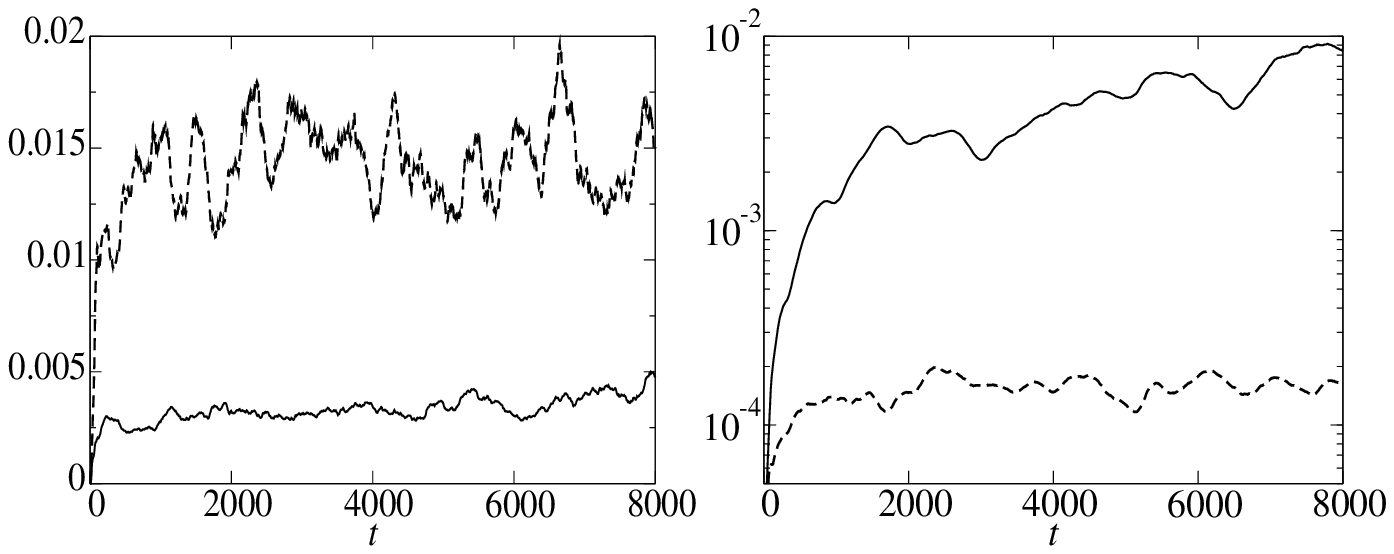}
}
\caption{Left: time evolution of the parallel dissipation $D_\|$ for the kinetic (run C, solid line) 
and magnetic (run D, dashed line) drivings in the conditions described in Section 3.2. 
Right: same for the perpendicular dissipation $D_\perp$  (in lin-log scale).}
\label{diss_par_smallbox}
\end{figure}

The contrast between the two types of forcing is even more striking at the level of the 
dissipations (fig. \ref{diss_par_smallbox}). While for magnetic driving, 
dissipation is strongly dominant on the parallel magnetic field fluctuations, with the kinetic one it 
mostly affects the transverse components, indicating a reduced transfer to sonic waves 
and the possibility of a direct cascade of Alfv\'enic modes,  in qualitative agreement with 
the weak-turbulence analysis performed by \citet{YF08} on the Vlasov equation.
Furthermore, when the kinetic forcing is monitored in order to maintain a prescribed energy 
as in Section 3.1, no energy gap establishes, indicating that the structures can sustain a 
much more constant direct transfer than in the case of Section 3.1. 
This suggests the possibility of a more standard turbulence when the injection
scale is sufficiently close to that at which structures form
and dispersion acts efficiently. 

The significant difference between the two forcings can be clarified by 
looking at the distribution of energy among the various 
modes and the different scales, presented  in fig.  \ref{figure_color}.
At large scale ($k=4$, panel m),  a  peak 
is visible at the ion cyclotron frequency for both drivings,
while whistler  modes are  also present  in the case of magnetic driving (panel o). 
The relative intensity of the magnetic versus kinetic components is consistent with the 
properties of the corresponding eigenmodes. Note that with a magnetic driving, the peaks are
especially sharp for the transverse modes. When the wavenumber is increased ($k= 25$), 
the peaks broaden under the effect of wave coupling.
This leads to  dominant ion cyclotron turbulence in the case of  a kinetic forcing (panel n).
With a magnetic driving (panel p),
the sonic mode, visible on the left of the whistler peak at large scale, becomes
dominant, while the whistler peak still remains conspicuous.
To summarize, the pressure balance observed in the case of a kinetic driving is consistent 
with the small amplitude of the intermediate modes, while the transverse dynamics is dominated 
by ion-cyclotron modes that, with the present forcing, can cascade down to the dissipation scales.  
Differently, in the case of a magnetic driving, the transverse velocity is dominated by 
ion-cyclotron modes and the transverse magnetic field to whistler modes.
The large amplitude of the latter waves prevents in particular the establishment of a pressure balance 
when the system is magnetically driven, which leads to an efficient generation of sonic modes that 
cascade and dissipate. In contrast, kinetic forcing does not drive whistler waves, 
so pressure balance can establish and sonic turbulence remains subdominant,
leaving the possibility for a direct cascade of ion-cyclotron modes.

\section{Oblique dynamics}

As already mentioned, for oblique propagation,
the system is driven on the $z$ component of the velocity or the magnetic field,
thus perpendicularly to the plane
defined by the ambient field and the direction of propagation.
The injection rate  is the same as the injection
on each component in the case of parallel propagation, and thus globally reduced by a factor two.
Two typical angles are considered, $\theta=45^0$ to exemplify a generic oblique regime,
and $\theta =80^0$ that corresponds to a quasi-transverse dynamics.

\subsection{The case of $45^0$ propagation angle}

With a kinetic driving, the same computational domains and forcing scales as in Sections 3.1 and 3.2
were considered, while in the case of a magnetic driving, only the largest domain was considered.
A resolution of $512$ mesh points was used in all cases, with
$\mu_x= \mu_y = \kappa_y = 2 \cdot 10^{-2}$, $\mu_z = \kappa_z  = 10^{-6}$ and
$\Delta t= 10^{-4}$ (large-scale forcing),
$\mu_x= \mu_y = \kappa_y = 5 \cdot 10^{-3}$, $\mu_z = \kappa_z  = 10^{-5}$ and
$\Delta t= 5 \cdot 10^{-5}$ (smaller-scale forcing).

\begin{figure}
\centerline{
\includegraphics[height=3.7cm,width=8cm]{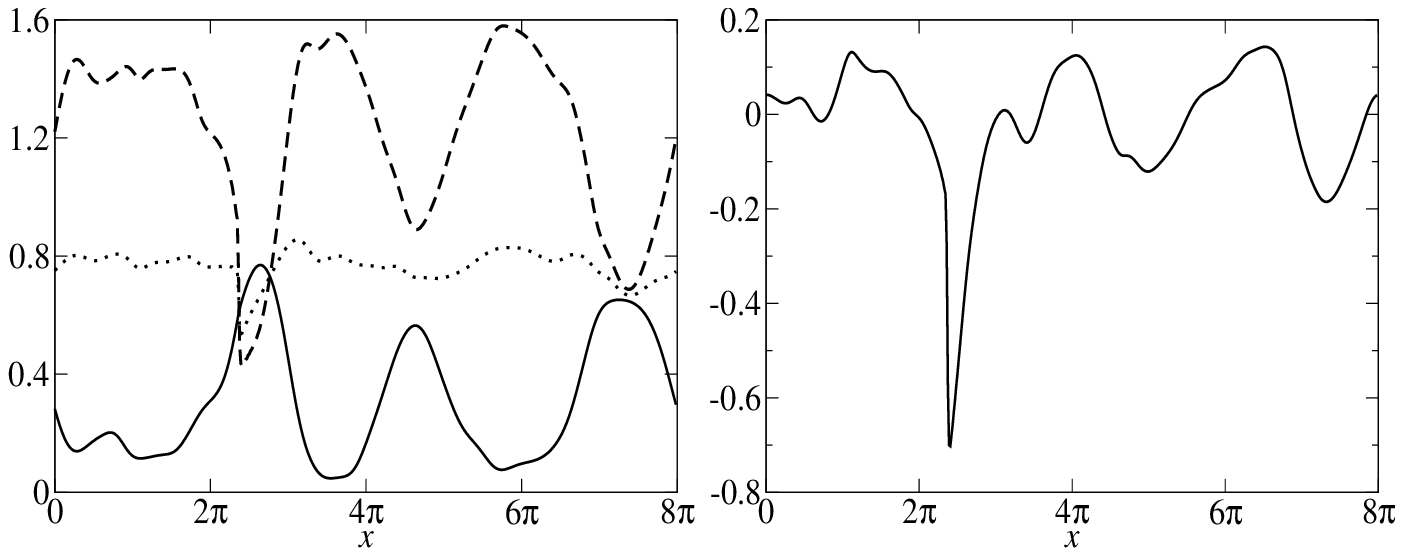}
}
\caption{Left: profile of the  the magnetic $|{\bf b}_\perp|^2/2$ (solid line)
and thermal $\beta \rho^{\gamma}/\gamma$ (dashed line) pressures, 
together with their half-sum (dotted line)
at time $t=422$, for the run E with large-scale kinetic driving described in Section 4.1.
Right: velocity component $v_x$ at the same time.}
\label{fields_45_obl_largebox}
\end{figure}

A main observation common to all these
simulations is the absence of a significant energy transfer to the largest scales, but rather the
establishment of a statistically steady state, with a total energy saturating at about the same
level in all the runs, after about a  thousand time units. In the case of the  kinetic driving
at large scales (run E), a pressure balance nevertheless establishes,
perturbed by localized events associated with the formation of cusps on the longitudinal
velocity profile (fig. \ref{fields_45_obl_largebox}). These peculiar
structures are reminiscent of the intermediate shocks that develop
in the Cohen-Kulsrud equation \citep{CK74}.
The pressure balance deteriorates when the driving is at smaller scales,
especially when it is of magnetic type.

At the level of the dissipations, the various types of driving do not induce qualitative
differences. It is noticeable that the dissipation
$D_z=\int (\mu_z(\partial_x v_z)^2 +  \kappa_z(\partial_x b_z)^2) dx $
affecting the velocity and magnetic field components in the transverse direction is smaller by a
factor of order $10^{4}$ than the dissipation
$D_{\rm plane}=\int (\mu_x(\partial_x v_x)^2 +  \mu_y(\partial_x v_y)^2 + \kappa_y(\partial_x b_y)^2) dx $
of the components in the (${\bf B}_0, \widehat{x}$) plane.

A more detailed understanding of the dynamics is provided by the frequency analysis displayed in
fig. \ref{figure_color}. With a large-scale kinetic forcing (run E: panels g, h, i), three
distinct frequencies corresponding to the three linear eigenmodes with comparable amplitudes,
are identified on the $k=1$ mode.
At smaller scales ($k=4$ and  $k=25$), in contrast, a peak near the intermediate frequency is dominant and
strongly broadened, with most of the energy contained in the component parallel to the direction
of propagation, consistent with the properties of the corresponding eigenmodes
displayed in fig. \ref{figure_linearmodes} (left). This confirms the compressible nature of
the turbulence that develops at small scales.

When kinetic energy is injected at smaller scales (run F: panels q, r), 
the three eigenfrequencies are still visible at
$k=4$, with a much more intense contribution of the whistler modes when compared
with the large-scale driving.
The presence of whistler at large or intermediate scales which is even more conspicuous  
in the case of magnetic driving (run G: panels s, t), is expected to be at the origin 
of the disruption of the pressure balance,
as also observed in the case of parallel propagation.
At smaller scale ($k=25$), the intermediate mode strongly dominates, as for the large-scale driving.
In other words, in oblique directions the presence of dispersion makes the dynamics fully compressible, 
in contrast with parallel propagation where an incompressible Alfv\'enic turbulence 
can be isolated when concentrating on the transverse components of the fields.

\subsection{The case of $80^o$ propagation angle}

\begin{figure}
\centerline{
\includegraphics[height=3.7cm,width=8cm]{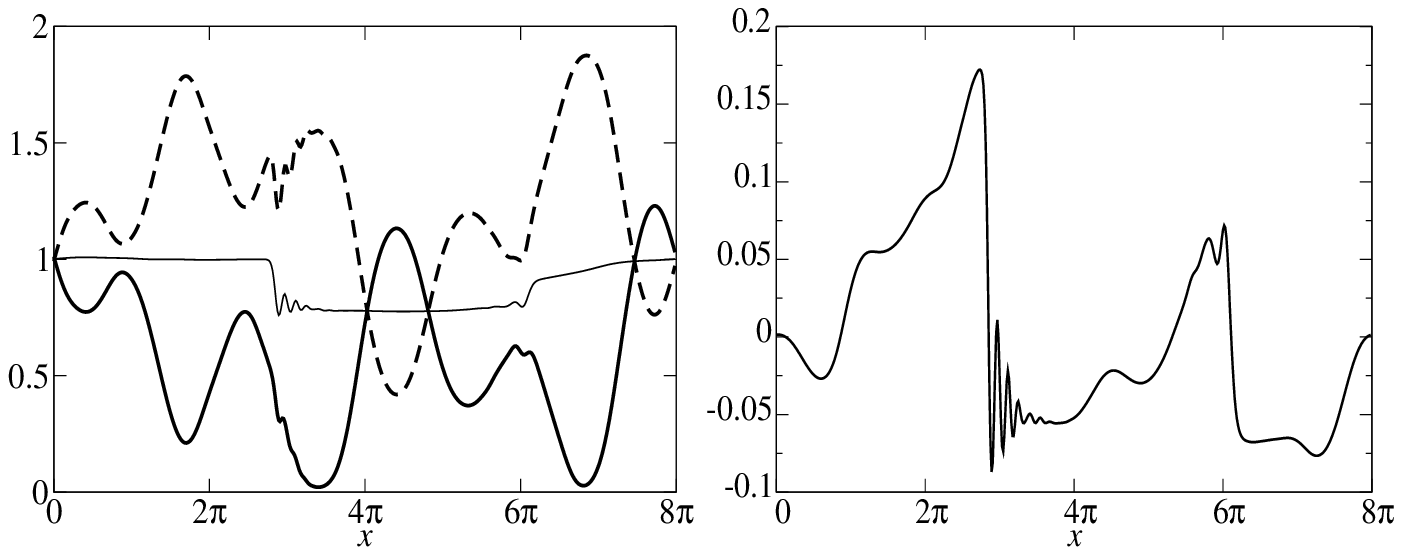}
}
\caption{Left: profile of the  the magnetic $|{\bf b}_\perp|^2/2$ (solid line)
and thermal $\beta \rho^{\gamma}/\gamma$ (dashed line) pressures, together with their half-sum (dotted line)
at time $t=2472$ for  run H with large-scale kinetic driving described in Section 4.2.
Right: velocity component $v_x$ at the same time.}
\label{fields_80_obl_largebox}
\end{figure}

\begin{figure}
\centerline{
\includegraphics[height=3.7cm,width=8cm]{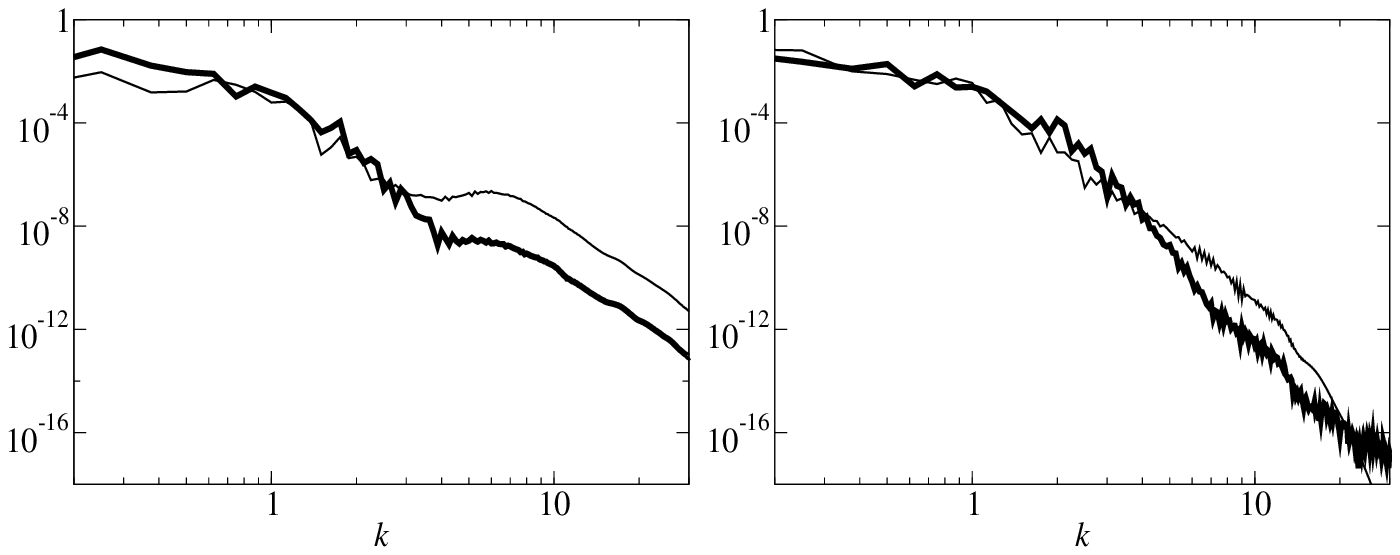}
}
\caption{Spectra $|\widehat{\bf v}_z|^2$ (thick line) and $|\widehat{\bf b}_z|^2$ (thin line) 
of the transverse fields, averaged over the time intervals $t=2465-24772477$ (left), and $t=10595-10565$ (right)
for run H.}
\label{spectra_80_largebox}
\end{figure}

In this case, the simulation was performed with a large-scale kinetic driving
(run H), in the same conditions as in Section 4.1.
A resolution of $512$ mesh points was used, with
$\mu_x= \mu_y = \kappa_y = 10^{-2}$, $\mu_z = \kappa_z  = 10^{-6}$ and $\Delta t= 10^{-4}$.
Compared with the $45^0$ case, the energy saturates at a much later time ($t \approx 40 000$).
This long time interval is required because of the presence of an early inverse cascade
that leads to a dominance of the $n=1$ mode, which nevertheless saturates
before the end of the simulation.
A pressure-balanced state establishes very rapidly (fig. \ref{fields_80_obl_largebox}, left).
It is nevertheless perturbed by dispersive shocks,
which occur very frequently and are especially visible on the velocity component $v_x$,
as displayed in fig. \ref{fields_80_obl_largebox} (right) at $t=2472$.
The spectra of the transverse velocity and magnetic fields are displayed in fig. \ref{spectra_80_largebox},
averaged over the time intervals $t=2465-2477$ (left) and $t=10595-10565$ (right).
Left panel represents a typical situation in which a dispersive shock is crossing the simulation box,
its spectral signature being conspicuous as a flat zone at intermediate scales.
Right panel represents instead the same spectra at a moment when no shocks are present.
In such a case, the velocity spectrum is steeper than the magnetic
one, in contrast with the case of parallel propagation where the
opposite takes place. This is consistent with the domination of
whistler modes at intermediate scales, as discussed below.

Inspection of fig. \ref{figure_color} (panels j, k, l) provides further insights on the dynamics.
Compared with the $45^o$ case (panels g, h, i), much more energy is contained
at low frequency in all the spatial Fourier modes.
The characteristics of the various field at large ($k=1$) and intermediate ($k=4$) scales where 
$v_y \approx v_z \gg v_x, b_y, b_z$,
indicate that these modes can be viewed as ion-cyclotron waves
(fig. \ref{figure_linearmodes}), which for this angle of propagation
identifies with the branch of kinetic Alfv\'en waves.
At smaller scales ($k=25$, panel l) $v_y \gg v_z$, probably because of the choice $\mu_y \gg \mu_z$. 
The oscillations in the dispersive shock displayed in fig. \ref{fields_80_obl_largebox} (right)
can be identified as the whistler modes associated with the broadened peak visible
on the $v_x$ field for $k=4$ displayed in panel k.
These high-frequency modes are indeed strongly dominated by the $v_x$ component at these scales,
which is consistent with the properties of the linear whistler waves.
As a consequence, they are not expected to significantly disrupt the pressure balance,
in contrast with the $45^o$ case with small-scale forcing. In the latter simulation, whistler modes
are indeed present at scales where they involve a significant magnetic contribution.
At $k=4$, a weak peak corresponding to the intermediate frequency is also present.
The corresponding  mode dominates over the whistler wave at $k=25$, where it reaches an amplitude
comparable to the low-frequency components (panel l).
The overall dynamics is then characterized by a complex distribution of energy among the different
waves, in contrast with the $45^o$ case where the dynamics is mostly governed by the intermediate modes.

Examination of the dissipations shows no difference between the $45^o$ and the $80^o$ cases for
what concerns $D_{\rm plane}$, while $D_z$ is in the mean larger by a factor 2  and displays
stronger fluctuations for propagation at $80^o$.
The presence of ion-cyclotron waves for quasi-transverse propagation without a major influence
on the dissipation $D_z$ (which is indeed comparable to that observed at $45^o$ where such modes are absent)
suggests that they are not able to cascade efficiently to the dissipation scales.

\section{Beyond the Hall-MHD description}

\begin{figure}
\centerline{
\includegraphics[height=3.7cm,width=8cm]{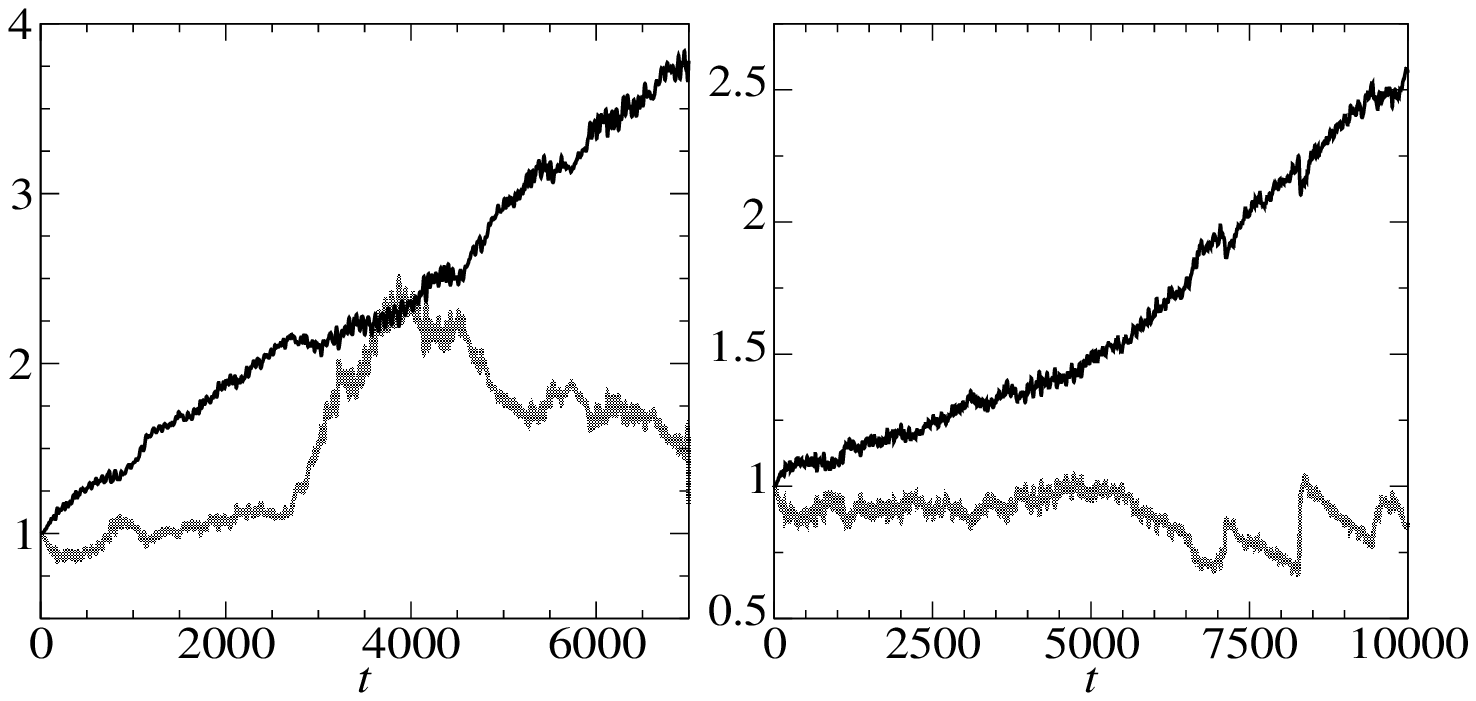}
}
\caption{Time evolution of the parallel (grey line) and transverse
  (black line) temperatures for simulations with $\beta_{\|i}=0.6$
(left) and  $\beta_{\|i}=1.2$ (right) for the runs described in Section 5.}
\label{temperatures}
\end{figure}

Although significantly richer than classical MHD, HMHD cannot capture important dynamical
properties of space plasmas originating from their quasi-collisionless character. Although such media
should require a fully kinetic description and thus enormous computational resources, it may be of interest to
deal with fluid models that extend MHD by retaining main
kinetic effects such as Landau damping and also finite Larmor
radius (FLR) corrections that, as stressed for example in
\citet{SCPVS07}, are expected to be comparable to the Hall effect within the magnetopause.
In spite of their semi-phenomenological nature, these so-called Landau fluid models correctly reproduce
the kinetic theory in the low-frequency limit.

First introduced in the context of large-scale MHD \citep{SHD97}, these  models
were then extended by retaining Hall effect and FLR corrections in order to describe quasi-transverse
ionic scales obeying the gyrokinetic scaling and thus associated with low frequencies. 
Nevertheless, in contrast with the gyrokinetic theory, the fast
magnetosonic modes are not averaged out in the Landau fluid approach
and their large-scale dynamics is accurately reproduced. A comprehensive derivation
of the model is given in \citet{PS07},
and its reduction to one space dimension can be found in \citet{BPS07}.

Preliminary simulations of the Landau fluid model, retaining Landau damping of the
ions and the electrons, together
with ion finite Larmor radius corrections, were performed in a domain of extension $16\pi$
for a propagation angle of $80^o$.
The Landau damping being relatively weak in quasi-transverse directions, 
we resorted to add a magnetic diffusivity ($\eta = 0.02$) together with a $k^8$-hyperviscosity $\nu_h = 10 ^{-8}$
in a simulation with a resolution of 256 collocation points.
The system is driven by a random forcing acting on the transverse components of the velocity field,
identical to that used in the HMHD simulation presented in section 4.2.
Nevertheless, in order to control the growth of the total energy that is
dominated by the global heating of the plasma, the forcing is monitored in order to maintain the
kinetic energy between prescribed upper and lower bounds.

As in the HMHD corresponding simulation, we observe the development of pressure-balanced structures
with the formation of small-scale fluctuations associated with whistler modes.
Compressible modes are nevertheless more damped than in HMHD simulations
due to Landau damping. An interesting issue that the Landau fluid can address
concerns the evolution of the plasma temperatures. Figure \ref{temperatures} displays the typical evolution
of the ion temperatures of a plasma with ion parallel beta $\beta_{\|i}$ equal to $0.6$ (left) and  $1.2$ (right).
We observe in both simulations that the
anisotropy progressively grows up, leading to dominant transverse ion
temperatures, making the plasma potentially unstable to ion mirror  instabilities. For
the electrons, the parallel temperature (not shown) increases, possibly leading to an electron firehose
instability.
Later on, we observe abrupt variations of parallel ion temperatures, resulting in  a reduction of
the anisotropy. These events are associated with the formation of quasi-singular structures on the
velocity components, that can be shown to affect more the parallel
than the perpendicular temperature.
The above temperature variations was checked to be due to mirror
instabilities, that is saturated thanks to the small-scale FLR
corrections retained by the model.
Differently, a phenomenological relaxation of temperature instabilities, 
in the form of effective collisions, was implemented in
\citet{SHQS06}. Such a model should be useful in situations
where ion-cyclotron or oblique firehose instabilities are excited,
as these instabilities cannot be described with fluid equations.

The Landau fluid simulations thus reproduce constraining effects due to temperature anisotropy instabilities
on the plasma parameters, also observed on solar wind data
\citep{HTK06}. As the plasma  anisotropy
relaxes under the effects of temperature instabilities,
the system enters a phase of slow dynamics. This regime is however not expected to
persist, as new temperature anisotropy can develop, making the above scenario to repeat.

\section{Conclusion}
Although limited to one space dimension, the present study reveals specific aspects of the turbulent dynamics
of magnetized plasmas at scales comparable to the ion inertial length,
in a regime where the transverse components of the velocity or the magnetic fields are randomly driven.
Special attention was paid to the distribution of the energy among the different MHD waves
that can be clearly identified from their linear dispersion relation, in spite of  a possible small 
shift in the temporal spectrum of the computed fields, 
in situations where the presence of large-scale coherent structures can be viewed as performing a 
renormalization of the ambient parameters.

A main observation in the case of parallel propagation is the
contrast between a large-scale forcing for which the energy is
almost entirely transferred to magnetosonic modes with nevertheless
a non-negligible transfer to larger scales,
and the regime where the driving takes place at scales where dispersion is more efficient,
for which, provided the driving is of kinetic type, a direct Alfv\'enic transfer establishes,  a result 
qualitatively consistent with the weak-turbulence analysis
performed by \citet{YF08} in the context of the Vlasov equation.
In oblique directions, the combined role of dispersion and compressibility leads to a
turbulence dominated by the intermediate modes, with an increasing contribution
of low frequency kinetic Alfv\'en waves, together with a faster establishment
of total pressure balance, as the propagation angle is increased. 
Furthermore, the Landau fluid model shades a light on the development of
pressure anisotropy resulting in recurrent instabilities.

To conclude, we would like to stress the complexity of the turbulence problem in magnetized
plasmas, even within the strongly simplified description provided by one-dimensional HMHD.
The usual picture of inertial ranges where energy ``cascades'' progressively from scale to scale
at a constant rate turns out to be strongly affected by the predominance of structures
and by a strongly fluctuating transfer, making possibly questionable the usual concepts of the
classical turbulence theory,  and leading to a dynamics that turns out to be significantly less universal.

\section*{Acknowledgments}
The work was supported by ``Programme National Terre Soleil'' of INSU-CNRS.

\bibliographystyle{elsarticle-harv}
\bibliography{biblio}

\end{document}